\newcommand{\eqb}{\begin{equation}}
\newcommand{\eqe}{\end{equation}}

\documentclass[aps, prb, letterpaper, amsmath, amssymb, notitlepage, twocolumn, showpacs]{revtex4}
\usepackage{graphicx}
\begin{document}

\title{Diffusion in a Half--Space: From Lord Kelvin to Path Integrals}

\author{Michael Slutsky}
\affiliation{Department of Physics, Massachusetts Institute of
Technology, 77 Massachusetts Avenue, Cambridge, Massachusetts 02139, USA}

\date{\today}

\begin{abstract}

Many important transport phenomena are described by simple
mathematical models rooted in the diffusion equation. Geometrical
constraints present in such phenomena often have influence of a
universal sort and manifest themselves in scaling relations and
stable distribution functions. In this paper, I present a
treatment of a random walk confined to a half--space using a
number of different approaches: diffusion equations, lattice
walks and path integrals. Potential generalizations are discussed
critically.

\end{abstract}

\pacs{05.40.Fb %Random walks and Levy flights
82.35.Lr %Physical properties of polymers
}

\maketitle

\section{Introduction}\label{sec:intro}

Diffusion is perhaps one of the most well--studied fields of
modern physics. Diffusion processes are ubiquitous and extremely
important in physics, biology, chemistry, material science and
other disciplines of basic and applied research. Many famous
names, such as Fourier, Kelvin and Einstein, are associated with
the theory of diffusion. About two hundred years after the
diffusion equation was first written down, the motivation for
studying diffusion by both theorists and experimentalists has not
faded. Modern research deals with questions of much greater
complexity than the originally formulated diffusion equation.
Fields such as pattern formation, non-equilibrium growth,
reaction-diffusion processes and granular media keep challenging
physicists and applied mathematicians and yield many interesting
and often unexpected results.

Compared to these actively developing fields, "old--fashioned"
linear diffusion may seem boring and trivial. Often it appears
though, that newly discovered phenomena and incessantly
developing techniques provide formal and conceptual contexts in
which old and well--known questions appear in a new light and
acquire additional scientific and pedagogical value. For
instance, rapid progress in the physics of polymers and critical
phenomena in the late 1960's and 1970's, stimulated by
introduction of field--theoretic ideas and methods, has shifted
the accents in  diffusion studies towards scaling and
universality. Many systems that appeared completely unrelated at
first glance turned out to exhibit identical behavior in certain
asymptotic (``critical'') regions of the parameter space.

In this paper, we study the rather old problem of a random walk
\footnote{A more concise term is {\em random flight}; we will
use the first one as more widely known, since the difference
between the two is inessential throughout this paper.}  confined
to a half--space, using a number of different approaches. First,
we review the traditional methods -- the diffusion equation and
lattice walks. Next, we discuss the applications of these methods
to Gaussian polymer chains, with emphasis on scaling. Finally, we
show how the same problem can be solved using basic
field--theoretic methods, i.e. functional integration in Fourier
space.  We shall see how important scaling relations and
universal distribution functions can be obtained in each case
through appropriate limiting procedures. We also discuss possible
ways to extend and generalize the problem for more realistic
systems.

\section{Diffusion and random walks}

\subsection{Boundary conditions}

Consider a $N$--step random walk starting at ${\bf r}_0 = 0$ in
the three--dimensional (3D) space.  Let $G({\bf r}, N)$ be the
probability density for the walk to end at ${\bf r}$.  It is well
known that for large $N$ and in the absence of obstacles and
boundaries, $G({\bf r}, N)$ is a solution of the {\em diffusion
equation}
\eqb\label{eq:diff}
(\partial_N - \nabla^2) G({\bf r}, N) 
= \delta({\bf r})\delta(N), 
\eqe
namely,
\eqb
G({\bf r}, N) = \frac{e^{-{\bf r}^2/(2 N)}}{(2\pi  N)^{3/2}}.
\eqe
It is reasonable therefore to assume that solutions of the same
kind can as well be found for any bounded region ${\mathbb R}$.

Naturally, one has to specify the boundary conditions. This is
not as trivial as it appears and, in fact, depends on the physical
context of the problem. If, for example, the random walker is
allowed to touch the boundary and then step back with probability
one, the {\em reflecting} boundary conditions are
appropriate. Formally, it means that the flux across the boundary
vanishes
\eqb
\nabla G({\bf r}, N) \cdot {\bf n} |_{{\bf r}\in\partial {\mathbb R}} = 0.
\eqe
Here ${\partial {\mathbb R}}$ denotes the boundary of the region
${\mathbb R}$ and ${\bf n}$ is a unit vector locally normal to
${\partial {\mathbb R}}$. Another possible choice of boundary
conditions corresponds to the case when the walker sticks to the
boundary upon reaching it -- the {\em absorbing} boundary
conditions
\eqb
G({\bf r}, N)|_{{\bf r}\in\partial{\mathbb R}} = 0.
\eqe
For reasons that become clear below, we shall mainly be
interested with the latter case.

\subsection{Method of images}

Consider a random walk starting at ${\bf r}_0 = \hat{\bf z}a$
away from the plane $z = 0$ and confined to the $z>0$
half--space. For the absorbing boundary, we expect the
probability distribution for the end point to satisfy the
following boundary value problem
\eqb\label{eq:diff_half}
\left\{
\begin{array}{l}
(\partial_N - \nabla^2)G({\bf r}, N; a) 
= \delta({\bf r} - \hat{\bf z}a)\delta(N) \\
G(z = 0) = 0,  \qquad G(|{\bf r}| \to \infty) = 0
\end{array}
\right.
\eqe
Any introductory textbook on PDEs contains a straightforward
solution of this problem, which consists of introducing a sink,
or {\em negative image}, at $(0, 0, -a)$ and extending the
problem to the entire space\footnote{It is straightforward to
verify that reflecting boundary conditions correspond to a
positive image.}. The boundary condition at $z=0$ is then
authomatically satisfied and the solution is
\eqb
G({\bf r}, N; a) = \frac{1}{(2\pi  N)^{3/2}}\left[
e^{-({\bf r }- \hat{\bf z}a)^2/(2 N)}
- e^{-({\bf r }+ \hat{\bf z}a)^2/(2 N)}
\right].
\eqe
For $a\ll\sqrt{N}$, we can expand the expression in
parentheses to obtain
\eqb
G({\bf r}, N; a) \simeq \left(\frac{2az}{N}\right)\frac{e^{-{\bf
r }^2/(2 N)}}{(2\pi N)^{3/2}}.
\eqe
We see that the probability distribution can be factorized into
$z$--dependent and $z$--independent parts. The latter, which
includes degrees of freedom parallel to the boundary, is not
affected by the presence of the boundary. Thus, in what follows
we will be predominantly occupied with the $z$--dependent part of
$G({\bf r}, N; a)$.

\subsection{A short historical digression}

The theory of diffusion was first developed in the beginning
of the 19th century by Joseph Fourier; his work was summarized in
the famous {\em Th\'eorie analytique de la
chaleur}~\cite{chaleur}, first published in 1822. It contains an
extensive treatment of {\em homogenous} heat diffusion problems
for a variety of geometries, mostly by the variable separation
method.

The first generalized approach to solving non-homogenous
diffusion probems was formulated by Sir William
Thomson~\cite{kelvin1}, more widely known as Lord Kelvin, in
1850. He realized that particular solutions can be obtained by
superposition of solutions for ``instantaneous simple point
sources'' (which are now called by physicists ``Dirac's
delta--functions''). In short, what he did was to invent the
Green's function method for the diffusion equation; it was later
used by E. W. Hobson to treat heat-conduction problems with a
variety of sources and boundary conditions~\cite{hobson}.

Kelvin was also the first to apply the method of images to
account for boundary conditions for electricity conduction
in a semi--infinite telegraph line~\cite{kelvin1}.

\subsection{``Phantom'' polymers and scaling}

As mentioned in the Introduction, random walks are often used to
model long polymer molecules. Proper description includes a
nontrivial requirement: the random walk must be self--avoiding,
i.e. no point in space should be visited more than once. This
requirement introduces long--range correlations and makes the
problem tractable only approximately. However, if the
self--avoidance is removed, the problem becomes much
simpler. Such a polymer is called {\em phantom}; alternative
names are {\em ideal} or {\em Gaussian} chain. The distribution
function for the end--to--end radius vector ${\bf r}$ of a
phantom polymer of length $N$ obeys the diffusion equation
(\ref{eq:diff}).

Rapid developments in polymer physics and the theory of critical
phenomena have revealed a number of {\em universal} properties
that arise in all polymer chains beyond a certain level of
coarse--graining~\cite{degennes_book,doiedwards}. These
properties are characterized by {\em scaling relations}. Perhaps
the most widely known scaling law relates the mean end--to--end
distance $R$ of a polymer chain to its length: $R\propto
N^\nu$. The number $1/\nu$ thus plays the role of the polymer's
fractal dimension. It is universal in that it depends only on the
dimensionality of the embedding space, e.g. in 3D, $\nu\approx
0.59$ for a self--avoiding polymer and $\nu = 1/2$ for a phantom
one. Another important scaling relation describes the number of
different configurations ${\mathfrak N}$ of a polymer
\begin{equation}\label{eq:gammadef}
{\mathfrak N}= {\rm const}\times \zeta^N N^{\gamma-1},
\end{equation}
where $\zeta$ is the ``effective coordination number'' that
depends on the microsopic details
(cf. Eq.~\ref{eq:conf_lattice}), and $\gamma$ is a universal
exponent.  The factor $\zeta^N$ can be thought of as counting the
configurations of an unconstrained $N$--step random walk with
$\zeta$ options available at each step, whereas $N^{\gamma-1}$
accounts for constraints such as self--avoidance, obstacles
present etc.

If the distribution function $G({\bf r}, N)$ is known, the
exponent $\gamma$ can be obtained in a very straightforward way,
by simply integrating $G({\bf r}, N)$ over the whole space.
Thus, a phantom polymer has $\gamma = 1$. It turns out that
incorporating self--avoidance constraint leads to $\gamma =
1.16$\footnote{The exact form of $G({\bf r}, N)$ for a
self--avoiding polymer is not known, so $\gamma$ is calculated
either numerically, or by renormalization group methods}. This
can be interpreted as the enhancement of available space for
a self--avoiding polymer which appears ``swollen'' compared to a
phantom one.

Consider a phantom polymer anchored to the $xy$ plane and
confined to the $z>0$ half--space. Assuming that the polymer is
strongly repelled from the $z = 0$ plane, what is the
corresponding boundary value problem?  Intuitively, it would seem
that we must adopt reflecting (zero flux) boundary conditions at
$z = 0$. However, a more careful analysis shows that this is
incorrect. Note that $G({\bf r}, N)$ is proportional to the
number of $N$--step paths of length $N$ ending at ${\bf
r}$. Since the plane is repelling, while counting the paths
contributing to $G({\bf r}, N)$, we should discard those touching
the $z = 0$ plane at least once. Thus, the {\em probability} and
not the {\em probability flux} should vanish at the boundary,
which corresponds to the absorbing boundary conditions. This is
quite counter--intuitive: while a polymer, being an entire path,
is repelled from the boundary, a fictitious random walker that we
employ to model this path is absorbed there!

Chandrasekhar~\cite{chandra} in his classic paper of 1943,
considered both kinds of boundary conditions (we will touch on
his derivation below). However, the abovementioned subtlety was
overlooked by a number of subsequent authors which lead to
incorrect calculation of average quantities in polymer adsorption
problems (for a very lucid discussion of this topic see the paper
by DiMarzio~\cite{dimarzio}).

Thus, the anchored Gaussian chain should be described by
Eq. (\ref{eq:diff_half}).  Apparently, we should take $a = 0^+$,
which leads to
\eqb
G({\bf r}, N; 0^+) = 0.
\eqe
Of course, this result does not make sense. To obtain the correct
result, we should have normalized the distribution function for a
finite value of $a$ and only then take the limit $a\to 0^+$,
which would yield
\eqb
G({\bf r}, N; 0^+) = \left(\frac{z}{N}\right)\frac{e^{-{\bf
r }^2/(2 N)}}{(2\pi N)}.
\eqe
However, normalization eliminated any information connected with
$\gamma$; indeed, any normalized distribution function by
definition has $\int d^3{\bf r}~G({\bf r}, N)  = 1$. Again, one
can see that the problem here is that of the order of taking
limits. For any physical polymer $a$ is finite, at least of the
order of the smallest coarse-graining scale -- the {\em persistence
length}. For small but finite $a$,
\eqb
{\mathfrak N} \propto  \int_{-\infty}^\infty dx~dy\int_0^\infty dz 
~\frac{az}{N^{5/2}}e^{-{\bf r}^2/2N} \propto aN^{-1/2}.
\eqe
From here, we can read off the value of $\gamma$ which for the
anchored random walk is denoted $\gamma_1$~\cite{debell}:
\eqb
\gamma_1 = \frac{1}{2}.
\eqe
Thus, we see that the presence of the plane reduces the number of
accessible configurations compared to the unconstrained case
which manifests itself in the scaling exponent $\gamma_1$.  Note
that reflecting boundary conditions produce $\gamma = 1$ so that
the number ${\mathfrak N}$ is unchanged relative to the
unconstrained case. This is yet another argument for
incorrectness of reflecting boundary conditions in this problem.

\subsection{Counting walks on a lattice}

Chandrasekhar~\cite{chandra} suggested a direct way of counting
the paths on a lattice when a reflecting or an absorbing boundary
is present. We will briefly describe the derivation for an
absorbing boundary. The reader is encouraged to read the original
paper which despite being written more than half a century ago,
remains one of the best introductions into random walks and
stochastic processes in general.

Consider a one--dimensional random walker on a lattice (discrete
$z$--axis) with absorbing boundary at $z=0$. Suppose, the walk
starts some distance $n$ from the origin; our task is to
calculate the number of paths leading from $n$ to some other
point $m$, {\em without touching} the boundary. It turns out
that it is easier to calculate the number of paths that do touch
the boundary and then to subtract it from the total number of
paths leading from $n$ to $m$. To do so, we make use of a very
elegant theorem -- {\em the reflection principle}. 

%%%%%%%%%%%%%%%%%%%%%%%%%%%%%%%%%%%%%%%%%%%%%%%%%%%%%%%%%%%%%%%%%%%%
\begin{figure}[htb]
\includegraphics[width=7cm]{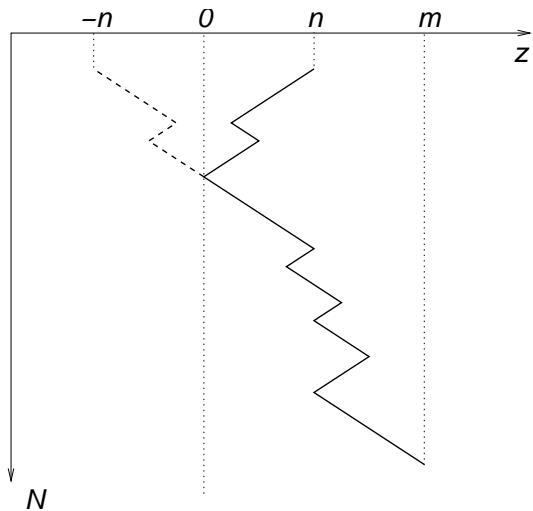}
\caption{\label{fig:reflection} The reflection principle.
}
\end{figure}
%%%%%%%%%%%%%%%%%%%%%%%%%%%%%%%%%%%%%%%%%%%%%%%%%%%%%%%%%%%%%%%%%%%%%

Let us extend our lattice to include the negative part of the
$z$--axis as well. Then, the reflection principle states that the
number of $N$--step paths originating at $n$, ending at $m$ and
touching or crossing the boundary $z = 0$ is equal to the number
of $N$--step paths that originate at $-n$ and end at $m$. Figure
\ref{fig:reflection} illustrates the reflection principle by presenting a way to
build a one--to--one mapping between the two sets of paths. Thus,
the number of paths not touching the boundary is
\eqb
{\mathfrak N} = 
\left(
\begin{array}{c}
N\\
\frac{1}{2}[N + m - n]
\end{array}
\right) - 
\left(
\begin{array}{c}
N\\
\frac{1}{2}[N + m + n]
\end{array}
\right).
\eqe
For the starting point near the boundary, and $m \ll N$, we can
expand the binomial coefficients using Stirling's formula to
obtain
\eqb\label{eq:conf_lattice}
{\mathfrak N} \simeq 2^{N}\left(\frac{2}{\pi N}\right)^{1/2}
\frac{m~e^{-m^2/(2N)}}{N}.
\eqe
Dividing by the total number of paths of length $N$ (which is
$2^{N}$), we obtain the probability density for a path to start
near the boundary and to end at some point $m$ without returning
to the boundary
\eqb
G(m, N) \simeq \frac{2m}{N} \frac{e^{-m^2/(2N)}}{(2\pi N)^{1/2}}.
\eqe

\section{Results from path integrals}

In this section, we employ the methods of functional (path)
integration to build a field--theoretical model that describes a
Gaussian chain anchored to an impenetrable plane.

\subsection{Example: unconstrained Gaussian chain}

In the field--theoretic approach, a flexible chain is described
by a function (``path'') ${\bf c}(\tau)$, where $\tau$ measures the position
along the chain.  The energy of a self--avoiding chain in an
external potential is given by~\cite{doiedwards,freed_book}:
\begin{eqnarray}
H[{\bf c}] = \frac{1}{2}\int_0^N \dot{\bf c}^2(\tau) ~d\tau + \int_0^N
U[{\bf c}(\tau)]d\tau
\nonumber \\
+ \frac{v}{2}\int_0^N \int_0^N\delta[{\bf c}(\tau) - {\bf
c}(\tau')] d\tau d\tau'.
\end{eqnarray}
The first two terms can be viewed as a harmonic potential between
neighboring segments of the chain and the external potential,
respectively, whereas the last one accounts for excluded
volume effects: each time the chain self--intersects, a penalty
of $v$ is paid. In what follows, we omit the self--avoidance
constraint.

The partition function for such a chain is a sum over all
possible paths ${\bf c}(\tau)$, given by a {\em path integral}
\eqb
\mathcal{Z}(N) = \int D[{\bf c}(\tau)]~e ^ {- H[{\bf c}]}.
\eqe
If the external potential is set to zero, this expression simply
counts the number of configurations of a Gaussian chain of length
$N$. The configurations are weighted with a weight $e ^ {- H[{\bf
c}]}$. If we want to count only paths starting at
the origin and leading to some point ${\bf r}$ then, after
normalization by $\mathcal{Z}(N)$, we obtain the probability
density
\eqb\label{eq:distr_pi}
G({\bf r}, N) = \frac{1}{\mathcal{Z}(N)}
\int D[{\bf c}(\tau)]~\delta[{\bf c}(N) - {\bf c}(0) - {\bf r}] e ^ {- H[{\bf c}]}.
\eqe
To calculate this sum, one has to define a measure of
integration. One way would be to discretize the chain and view
$\mathcal{Z}(N)$ as a limit of a multidimensional integral. In
this case, the problem is almost identical to calculating a free
particle propagator {\em a l\'a} Feynman \&
Hibbs~\cite{feynman_hibbs}. Another way around is to count the
Fourier--components of ${\bf c}(\tau)$; this way is somewhat
easier, since for harmonic Hamiltonians, degrees of freedom
decouple in Fourier space. Thus, setting
\eqb
\int D[{\bf c}(\tau)] \to \int \prod_{q}\frac{d^3\tilde{\bf c}(q)}{(2\pi)^3}
\eqe
and Fourier--transforming Equation (\ref{eq:distr_pi}), we obtain
%\begin{widetext}
\begin{align}
G({\bf k}, N) = \frac{\displaystyle
\prod_{q}\int\frac{d^3\tilde{\bf c}(q)}{(2\pi)^3}
~e^{i{\bf k}\cdot\tilde{\bf c}(q)[e^{iqN} - 1]
 - \frac{1}{2}q^2 |\tilde{\bf c}(q)|^2
}
}
{\displaystyle
\prod_{q}\int\frac{d^3\tilde{\bf c}(q)}{(2\pi)^3}
~e^{- \frac{1}{2}q^2|\tilde{\bf c}(q)|^2
}
}
\end{align}
%\end{widetext}
Both the numerator and the denominator contain products of
Gaussian integrals that can be calculated by ``completing the
square.'' The result is
\eqb
G({\bf k}, N) = \exp \left( 
- {{\bf k}^2}\int_{-\infty}^{+\infty} \frac{dq}{2\pi}\frac{1-\cos qN}{q^2}
\right) = e^{-{\bf k^2} N/2},
\eqe
which is the Fourier--transform of 
\eqb
G({\bf r}, N) = \frac{e^{-{\bf r}^2/(2 N)}}{(2\pi  N)^{3/2}}.
\eqe

\subsection{Anchored polymer - the partition function}
Now that we are somewhat familiar with the methodology of path
integrals, we return to the original problem -- the anchored
Gaussian chain in a half--space $z>0$. Writing it in terms of
path integrals, we immediately see that our task will not be as
simple as before. The reason for
this is that now possible values of $c_z(\tau)$ should be
positive. If we stay in  real space and calculate the limit of
a multidimensional integral, we see that ``completing the
square'' does not work because of this constraint, since the
resulting integrals cannot be calculated analytically. If we
decide to move to  Fourier space, it is not even clear how to
define the measure of integration.

The suggested way out of this complication is as follows. We
allow the polymer to cross the boundary and introduce a strong
repulsive interaction between the plane at $z=0$ and the chain.
Each time the polymer crosses or touches the plane, it is
``penalized'' by a large amount of energy. The modified
Hamiltonian is then
\eqb
H = H_0 + H_1,
\eqe
where
\eqb
H_0 = \frac{1}{2}\int_0^N \dot{\bf c}^2 ~d\tau,
\eqe
and
\eqb
H_1 = g\int_0^N\delta[c_z(\tau) - c_z(0)]~d\tau.
\eqe
 Thus, we expect
that when the coupling constant $g > 0$ becomes infinitely large, the
polymer will be entirely on one (either positive or negative)
side of the plane. The partition function is then
\eqb\label{eq:pf}
\mathcal{Z}(g,N) = \int D[{\bf c}(\tau)]~e ^ {- H_0[{\bf c}] - H_1[g, {\bf c}]}.
\eqe

To evaluate $\mathcal{Z}(g,N)$, we expand the integrand in
Eq.~(\ref{eq:pf}) in powers of $g$. Such an expansion could be
problematic when $g$ is large, and this is the limit we are
primarily interested in. However, if we are able to calculate the
general term of the expansion and to perform the summation to
infinity, this approach is valid.

The $n$-th ($n = 1,2,...$) term of the expansion reads
\eqb\label{eq:term}
\frac{(-g)^n}{n!} \int D[{\bf c}(\tau)]~e ^ {- H_0[{\bf c}]}\prod_{l
= 1}^{n}\int~\delta[c_z(\tau_l) - c_z(0)]~d\tau_l.
\eqe
Ordering the set $\{\tau_l\}$ and Fourier-transforming the
$\delta$-functions, this can be rewritten as
\begin{widetext}
\eqb
\left(\frac{-g}{2\pi}\right)^n \int D[{\bf c}(\tau)] ~e ^ {- H_0[{\bf c}]}
\int_0^N d\tau_n \int_0^{\tau_n} d\tau_{n-1} ... \int_0^{\tau_2}
d\tau_{1}\prod_{l
= 1}^{n}\int_{-\infty}^{+\infty} dk_l~e^{-ik_l[c_z(\tau_l) - c_z(0)]}.
\eqe
Henceforth, we focus on the $c_z(\tau)$ and denote it $c(\tau)$ for the sake
of simplicity. Integrating it out by the method discussed above
(i.e. ``completing the square''), we are left with
\eqb
\left(\frac{-g}{2\pi}\right)^n 
\int_0^N d\tau_n \int_0^{\tau_n} d\tau_{n-1} ... \int_0^{\tau_2}
d\tau_{1}\int_{-\infty}^{+\infty}
dk_1...dk_l~\exp\left[-\frac{1}{2}\sum_{l,m}k_l
\mathbb{T}^{(n)}_{lm}k_m \right].
\eqe
%\end{widetext}
Here,
\eqb
\mathbb{T}^{(n)}_{lm} = \tau_{\rm{min}[l,m]} = 
\left(
\begin{array}{cccccc}
\tau_1 & \tau_1 & \tau_1 & \cdot & \cdot & \tau_1 \\
\tau_1 & \tau_2 & \tau_2 & \cdot & \cdot & \tau_2 \\
\tau_1 & \tau_2 & \tau_3 & \cdot & \cdot & \tau_3 \\
\cdot &  \cdot & \cdot & \cdot &  \cdot & \cdot \\ 
\cdot &  \cdot & \cdot & \cdot &  \cdot & \cdot \\ 
\tau_1 & \tau_2 & \tau_3 & \cdot & \cdot & \tau_n 
\end{array}
\right).
\eqe
To perform multiple integration over $k_i$ we use the well--known
formula
\eqb
\int_{-\infty}^{+\infty}
\prod_j dk_j~\exp\left(-\frac{1}{2}\sum_{l,m}k_l
\mathbb{T}^{(n)}_{lm}k_m \right) = 
\sqrt{\frac{\displaystyle (2\pi)^n}{\displaystyle \rm{det} \mathbb{T}^{(n)}}}.
\eqe
It is straightforward to show that
\eqb\label{eq:Tn}
\rm{det} \mathbb{T}^{(n)} =
\tau_1(\tau_2-\tau_1)(\tau_3-\tau_2)...(\tau_n - \tau_{n-1}),
\eqe
so that after integration over $k_i$, Eq.~(\ref{eq:term})
reduces to
%\begin{widetext}
\begin{align}
\lefteqn{\left(\frac{-g}{\sqrt{2\pi}}\right)^n 
\int_0^N d\tau_n 
\int_0^{\tau_n} \frac{d\tau_{n-1}}{\sqrt{\tau_n - \tau_{n-1}}} ...
\int_0^{\tau_3}\frac{d\tau_{2}}{\sqrt{\tau_3-\tau_2}} 
\int_0^{\tau_2}\frac{d\tau_{1}}{\sqrt{\tau_1(\tau_2-\tau_1)}}}
\nonumber \\
&&{} = \left(\frac{-g}{\sqrt{2\pi}}\right)^n 
\int_0^N d\tau_n \tau_n^{n/2 - 1}\prod_{m =
1}^{n-1}\int_0^1~x^{m/2 - 1}(1-x)^{-1/2}dx \nonumber \\
&&{} = \frac{\left(-g\sqrt{{N}/{2}}\right)^n}{\displaystyle\Gamma\left(\frac{n}{2}+1\right)}
\equiv \frac{(-\hat{g})^n}{\displaystyle\Gamma\left(\frac{n}{2}+1\right)},
\end{align}
\end{widetext}
where $\hat{g} \equiv g\sqrt{N/{2}}$. Thus,
\eqb
\mathcal{Z}(g, N) = \mathcal{Z}(\hat{g}) = \sum_{n = 0}^\infty
\frac{(-\hat{g})^n}{\displaystyle\Gamma\left(\frac{n}{2}+1\right)} =
e^{\hat{g}^2}[1 - \Phi(\hat{g})],
\eqe
where 
\eqb
\Phi(x) = \frac{2}{\sqrt{\pi}}\int_0^{x}e^{-t^2}dt
\eqe
is the error function.

When $N\to\infty$, so does $\hat{g}$; expanding $\mathcal{Z}(\hat{g})$
for large $\hat{g}$, we obtain
\eqb
\mathcal{Z}(\hat{g}) =
\frac{1}{\sqrt{\pi}}\left(\frac{1}{\hat{g}} - \frac{1}{2\hat{g}^3} + O(\hat{g}^{-5}).
\right).
\eqe
Hence,
\eqb
\gamma_1 = 1 + \lim_{N\to\infty} \frac{\partial \ln
\mathcal{Z}}{\partial \ln N} = \frac{1}{2},
\eqe
which is identical to the value we obtained in the previous
section. 

Several remarks should be made at this point. First, note that
while calculating the general term of the expansion, we omitted
the common factor
\eqb
\mathcal{Z}_0(N) = \int D[{\bf c}(\tau)]~e ^ {- H_0[{\bf c}]},
\eqe
which actually has the form $\zeta^N$ (c.f.
Eq. (\ref{eq:gammadef})). Secondly, the coupling constant $g$
plays here the role of the inverse cutoff length, which often
occurs in field theory. Finally, we note that for {\em any} value
of $g > 0$, we can find $N$ large enough to make the
non-dimensional coupling $\hat{g} = g\sqrt{N/2} \gg 1$, so that
the number of accessible configurations scales as $N^{-1/2}$
relative to the unconstrained case. This important observation is
a signature of $universality$: no matter how small $g$ is, for
long enough polymer the overall repulsion is infinitely strong!

\subsection{Probability distribution}

The Fourier--transform of the (unnormalized) probability
distribution function (PDF) for the end-to-end distance of the
random walk is given by

\eqb
G({\bf q}, N) = \int D[{\bf c}(\tau)]~e^{-i{\bf q}\cdot [{\bf
c}(N) - {\bf c}(0)]}~e ^ {- H_0[{\bf c}] - H_1[g, {\bf c}]}.
\eqe
As above, we focus only on the $z$-dependent part of the PDF
$G(z, N)$. Expanding the path integral in powers of $g$, and
integrating out $c(\tau)$ we observe that the $n$-th term of the
expansion reads
\begin{widetext}
\begin{align}\label{eq:pdf_four}
\lefteqn{\left(\frac{-g}{2\pi}\right)^n 
\int_0^N d\tau_n \int_0^{\tau_n} d\tau_{n-1} ... \int_0^{\tau_2}
d\tau_{1}\int \prod_l dk_l~
\exp\left[-\frac{1}{2}(q^2N + \sum_l k_l\tau_l + \sum_{l,m}k_l
\mathbb{T}^{(n)}_{lm}k_m) \right]}\nonumber \\
&& {}= \left(\frac{-g}{\sqrt{2\pi}}\right)^n 
\int_0^N d\tau_n \int_0^{\tau_n} d\tau_{n-1} ... \int_0^{\tau_2}
\frac{d\tau_{1}}{\sqrt{\rm{det} \mathbb{T}^{(n)}}}
\exp\left[-\frac{q^2}{2}(N - \sum_{l,m}\tau_l
[\mathbb{T}^{(n)}]^{-1}_{lm}\tau_m) \right],
\end{align}
%\end{widetext}
where $\mathbb{T}^{(n)}$ is given by Eq.~(\ref{eq:Tn}). Now,
%\begin{widetext}
\eqb
[\mathbb{T}^{(n)}]^{-1}_{lm} = 
\left\{
\begin{array}{lr}
-\frac{\displaystyle\delta_{l+1,m}}{\displaystyle\tau_{l+1}-\tau_l} -
\frac{\displaystyle\delta_{l-1,m}}{\displaystyle\tau_l-\tau_{l-1}} +
\delta_{m,l}\left(\frac{\displaystyle 1}{\displaystyle\tau_{l+1}-\tau_l}+
\frac{\displaystyle 1}{\displaystyle\tau_l-\tau_{l-1}}\right)
& \textrm{$l\neq n, 0$} \\
-\frac{\displaystyle\delta_{2,m}}{\displaystyle\tau_{2}-\tau_1} -
\delta_{1,l}\left(\frac{\displaystyle 1}{\displaystyle\tau_{2}-\tau_1}+
\frac{\displaystyle 1}{\displaystyle\tau_1}\right)
& \textrm{$l = 1$} \\
-\frac{\displaystyle\delta_{n-1,m}}{\displaystyle\tau_n-\tau_{n-1}} +
\frac{\displaystyle\delta_{m,n}}{\displaystyle\tau_n-\tau_{n-1}}
& \textrm{$l = n$}
\end{array}
\right.
\eqe

Straightforward calculation yields
\eqb
\sum_{l,m}\tau_l[\mathbb{T}^{(n)}]^{-1}_{lm}\tau_m = \tau_n.
\eqe
Equation~(\ref{eq:pdf_four}) therefore reduces to

\begin{align}
\lefteqn{\left(\frac{-g}{\sqrt{2\pi}}\right)^n 
\int_0^N d\tau_n \int_0^{\tau_n} d\tau_{n-1} ... \int_0^{\tau_2}
\frac{d\tau_{1}}{\sqrt{\rm{det} \mathbb{T}^{(n)}}}
\exp\left[-\frac{q^2}{2}(N - \tau_n) \right]} \nonumber \\
&&{}= \frac{(-g)^n~e^{-q^2N/2}}{2^{n/2}\Gamma(n/2)} 
\int_0^N \frac{d\tau}{\tau} \tau^{n/2}
e^{ q^2\tau/2}
= \frac{(-\hat{g})^n~e^{-q^2N/2}}{\Gamma(n/2)} 
\int_0^1 \frac{ds}{s} s^{n/2}
e^{(q^2N/2)s}.
\end{align}
\end{widetext}
Summing over $n$, we obtain
\begin{eqnarray}
\tilde{G}(q,N)
= e^{-q^2N/2} \left(1 + 2\int_0^1
\frac{ds}{s}e^{(q^2N/2)s^2}~f(\hat{g}s)\right),
\end{eqnarray}
where
\eqb
f(x) = -\frac{x}{\sqrt{\pi}} + x^2e^{x^2}(1 - \Phi(x)).
\eqe
Thus,
\eqb
{G}(z,N) = \frac{e^{-z^2/2N}}{\sqrt{2\pi N}} F(\hat{g},
\hat{z}),
\eqe
where $\hat{z} \equiv z/{\sqrt{N}}$ and
\eqb
F\left(\hat{g}, \hat{z}\right) = 1 + 2\int_0^1
\frac{ds}{s}
\frac{\displaystyle e^{-\frac{1}{2}\hat{z}^2s^2/(1-s^2)}}
{\displaystyle\sqrt{1-s^2}}~f(\hat{g}s)
\eqe
is the {\em scaling function}.  To calculate $F\left(\hat{g},
\hat{z}\right)$ we rewrite this expression as follows
%\begin{widetext}
\eqb\label{eq:scalfun}
F\left(\hat{g}, \hat{z}\right) = A(\hat{g}) - 2\int_0^1
\frac{ds}{s}
\frac{\displaystyle 1 - e^{-\frac{1}{2}\hat{z}^2s^2/(1-s^2)}}
{\displaystyle\sqrt{1-s^2}}~f(\hat{g}s),
\eqe
%\end{widetext}
where
\eqb
A(\hat{g}) = 1 + 2\int_0^1
\frac{ds}{s\sqrt{1-s^2}}~f(\hat{g}s).
\eqe

It seems that the integral in Eq.~(\ref{eq:scalfun}) cannot be
calculated analytically in general, i.e. for arbitrary $\hat{g}$
and $\hat{z}$. However, since we are interested in the limit
$\hat{g} \gg 1$, we can easily calculate the leading term. We
note that the integrand is essentially nonzero only for values of
$s$ larger than some value $s_0(\hat{z})$. However small
$s_0(\hat{z})$ is, we can always take $\hat{g}$ large enough to
make $\hat{g}s_0(\hat{z}) \gg 1$. Thus, we can take 
\eqb
f(\hat{g}s) \simeq -\frac{1}{2\sqrt{\pi}\hat{g}s},
\eqe
so that
\eqb
F\left(\hat{g}, \hat{z}\right) \simeq  A(\hat{g}) +
\frac{1}{\sqrt{\pi}\hat{g}} 
\int_0^1\frac{ds}{s^2}\frac{\displaystyle 1 - e^{-\frac{1}{2}\hat{z}^2s^2/(1-s^2)}}
{\displaystyle\sqrt{1-s^2}}.
\eqe
Using a substitution 
\eqb
u = \frac{|\hat{z}|s}{\sqrt{2(1-s^2)}},
\eqe
we finally obtain
\begin{align}
F\left(\hat{g}, \hat{z}\right) \simeq  A(\hat{g}) +
\frac{|\hat{z}|}{\sqrt{2\pi}\hat{g}}
\int_0^\infty\frac{du}{u^2}\left(1 - e^{-u^2}\right) \nonumber \\
= A(\hat{g})
+ \frac{|\hat{z}|}{\sqrt{2}\hat{g}}.
\end{align}
For large values of $\hat{g}$, we have
\eqb
A(\hat{g}) = \frac{1}{2\sqrt{\pi}\hat{g}^2} + O(\hat{g}^{-4}).
\eqe
Thus, when $\hat{g}\to\infty$, the scaling function is linear in
$\hat{z}$ and the normalized PDF has the form
\eqb
G(z,N) = \frac{|z|}{2N}e^{-z^2/2N}.
\eqe
Apart from a factor of 2, this function is identical to the one
obtained in the previous sections. This factor appears because now
the chain can be either in the $z<0$ or in the $z>0$ half--space.

%%%%%%%%%%%%%%%%%%%%%%%%%%%%%%%%%%%%%%%%%%%%%%%%%%%%%%%%%%%%%%%%%%%%
\begin{figure}[htb]
\includegraphics[width=9cm]{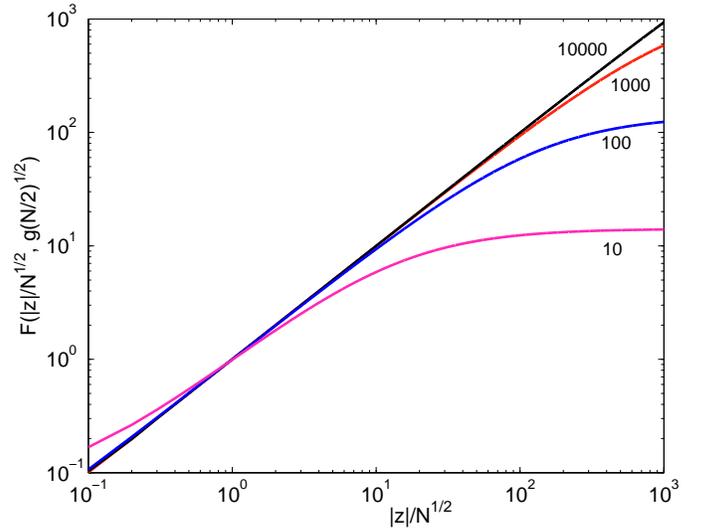}
\caption{\label{fig:scalfun} The (normalized) scaling function
$F\left(\hat{g},\hat{z}\right)$. Different curves are labeled by
corresponding values of $\hat{g}$.
}
\end{figure}
%%%%%%%%%%%%%%%%%%%%%%%%%%%%%%%%%%%%%%%%%%%%%%%%%%%%%%%%%%%%%%%%%%%%%

Figure \ref{fig:scalfun} shows the numerically computed scaling function
$F\left(\hat{g},\hat{z}\right)$ for different values of
$\hat{g}$. As expected, the larger $\hat{g}$ is, the
closer is $F\left(\hat{g},\hat{z}\right)$ to the linear
dependence.

\section{Conclusion}

To say the least, functional integration is not the most
effective way to obtain the probability distribution
$G(z,N)$. Why then has one to work so hard if it is possible to
obtain the answer in just a few lines?

Beside its clear educational value, path integral analysis of
random walks is a much more flexible and powerful tool when it
comes to real systems such as polymers in
solvents. The self--avoidance constraint that we omitted so readily
introduces long--range correlations that make the traditional
approaches loose their power and elegance. For example,
factorization of $G({\bf r},N)$ into transverse and longitudinal
parts is not valid anymore and integrating out one set of degrees
of freedom introduces non--local interactions into the other.
Whereas lattice walks are useful for Monte--Carlo simulations and
diffusion equations could perhaps be modified to include
mean--field corrections, field theory and primarily the
renormalization group (RG) is today the only analytical tool to
obtain universal scaling relations and phase diagrams using
controlled approximations~\cite{freed_book}. For instance,
generalizing our Hamiltonian to include self--interactions,
interactions between a number of polymers, external potentials
etc., it is possible to quantify the influence of space
dimensionality and learn about the relevance of various
interactions~\cite{douglas1,douglas2}.

\acknowledgments
I wish to thank Professor Mehran Kardar for many stimulating
discussions and Professor Martin Bazant and Yariv Kafri for
valuable advice. Special thanks are due Joe Levine for
important comments and thorough proofreading of the manuscript.

\end{document}